\documentstyle[aps,epsfig,prl,graphicx,twocolumn]{revtex}

\begin{document}
\title{Impact of the transport supercurrent on the Josephson effect}
\author{S.N. Shevchenko}
\address{B. Verkin Institute for Low Temperature Physics and Engineering, 47 Lenin\\
Ave., 61103, Kharkov, Ukraine.\\
E-mail: sshevchenko@ilt.kharkov.ua}
\maketitle

\begin{abstract}
We study the weak link between current-carrying superconductors, both
conventional and $d$-wave. The state of the system is controlled by two
parameters: the order parameter phase difference $\phi $ and the superfluid
velocity $v_{s}$, which parameterizes the parallel to the boundary transport
supercurrent which is injected externally. The low-temperature current-phase
relations are derived. We consider two models of weak links: a constriction
between two conventional superconductors and a plane boundary between two
differently orientated $d$-wave superconductors. We show that for some
relation between $\phi $ and $v_{s}$ quasiparticles create the current along
the boundary which flows in the direction opposite to the transport
supercurrent.
\end{abstract}


\section{Introduction}

It was demonstrated both experimentally \cite{Braunisch}-\cite{Ilichev} and
theoretically \cite{Higashitani},\cite{FRS97} that in some high-$T_{c}$
superconductors the paramagnetic Meissner effect takes place. Namely, at the
boundary of a {\it d}-wave superconductor placed in external magnetic field
the current flows in the direction opposite to the diamagnetic Meissner
supercurrent which screens the external magnetic field. This countercurrent
is carried by the surface-induced quasiparticle states. These nonthermal
quasiparticles appear because of the sign change of the order parameter
along the reflected quasiparticle trajectory. Such a depairing mechanism is
absent in the homogeneous situation. At $T=0$ in a homogeneous conventional
superconductor the quasiparticles appear only when the Landau criterion is
violated, at $v_{s}>\Delta _{0}/p_{F}$. (Here $v_{s}$\ is the superfluid
velocity which parameterizes the current-carrying state, $\Delta _{0}$
stands for the bulk order parameter, and $p_{F}$\ is the Fermi momentum.)

In a weak link of two superconductors with order parameter phase difference $%
\phi $, the order parameter can be tuned by $\phi $. In the constriction
between two conventional superconductors the proximity order parameter at
the contact is given by \cite{KO} $\Delta _{\phi }=\Delta _{0}\left| \cos
(\phi /2)\right| $. Suppose there is supercurrent tangential to the boundary
between two clean conventional superconductors. Namely, let us consider a
weak link between current-carrying superconductors. For some values of phase
difference $\phi $\ and superfluid velocity $v_{s}$, when the condition $%
v_{s}p_{F}>\Delta _{\phi }$ is satisfied, the quasiparticles appear in the
vicinity of the weak link. These quasiparticles create current in the
direction opposite to the transport supercurrent in the banks. The aim of
the present paper is to study this countercurrent theoretically.

First we consider a weak link between two conventional superconductors as a
constriction (see also \cite{K0Sh(2003)}), which is the simplest situation
to demonstrate how the countercurrent appears and how the interface-induced
quasiparticle states influence the current through the contact (Josephson
current). We then consider a weak link between two differently orientated
{\it d}-wave superconductors (see also \cite{KOSh(2004)}). The influence of
the transport supercurrent on the state of such a system is interesting in
the light of possible applications \cite{Jtransistor}, \cite{dd-qubit}.

\section{The model. General remarks}

We consider a perfect contact between two clean singlet superconductors. The
external order parameter phase difference $\phi $ is assumed to drop at the
contact plane at $x=0$. The homogeneous supercurrent flows in the banks of
the contact along the $y$-axis, parallel to the boundary. The sample is
assumed to be smaller than the London penetration depth so that the
externally injected transport supercurrent can indeed be treated as
homogeneous far from the weak link. The size of the weak link is assumed to
be smaller than the coherence length. Such a system can be quantitatively
described by the Eilenberger equation. Taking transport supercurrent into
account leads to the Doppler shift of the energy variable by ${\bf p}_{F}%
{\bf v}_{s}$. The standard procedure of matching the solutions of the bulk
Eilenberger equations at the boundary gives the Matsubara Green's function $%
\widehat{G}_{\omega }(0)$ at the contact at $x=0$ \cite{KOSh(2004)}. The
component $G_{\omega }^{11}(0)\equiv g_{\omega }(0)$ of $\widehat{G}_{\omega
}(0)$ defines current density at the boundary:
\begin{equation}
{\bf j}(0)=4\pi eN_{0}v_{F}T\sum\limits_{\omega _{n}>0}\left\langle \widehat{%
{\bf v}}%
\mathop{\rm Im}%
g_{\omega }(0)\right\rangle _{\widehat{{\bf v}}},  \label{j(0)}
\end{equation}
\begin{equation}
g_{\omega }(0)=\frac{\widetilde{\omega }(\Omega _{L}+\Omega
_{R})-isgn(v_{x})\Delta _{L}\Delta _{R}\sin \phi }{\Omega _{L}\Omega _{R}+%
\widetilde{\omega }^{2}+\Delta _{L}\Delta _{R}\cos \phi },  \label{g(0)}
\end{equation}
where $N_{0}$ is the density of states at the Fermi level, $\left\langle
...\right\rangle _{\widehat{{\bf v}}}$ denotes averaging over the directions
of Fermi velocity ${\bf v}_{F}$, $\widehat{{\bf v}}={\bf v}_{F}/v_{F}$ is
the unit vector in the direction of ${\bf v}_{F}$, $\omega _{n}=\pi T(2n+1)$
are Matsubara frequencies, $\Delta _{L,R}$ stands for the order parameter in
the left (right) bank, $\widetilde{\omega }=\omega _{n}+i{\bf p}_{F}{\bf v}%
_{s}$, $\Omega _{L,R}=\sqrt{\widetilde{\omega }^{2}+\Delta _{L,R}^{2}}$.

Analytic continuation of $g_{\omega }(0)$ ({\it i.e.} $\omega
_{n}\rightarrow -i\varepsilon +0$) gives the retarded Green's function; the
poles of $g_{\varepsilon }(0)$ determine the energy of the local Andreev
states in the system ({\it i.e. }bound states at the interface). The
direction-dependent Doppler shift ${\bf p}_{F}{\bf v}_{s}$ results in the
modification of current-phase dependencies and, particularly, in the
appearance of the countercurrent along the boundary. The following sections
discuss this phenomenon for the contacts of two conventional and of two {\it %
d}-wave superconductors.

\section{Weak link between current-carrying conventional superconductors}

In the case of a weak link in the form of a constriction between two
conventional superconductors with $\Delta _{L}=\Delta _{R}=\Delta _{0}$ ($%
\Delta _{0}=\Delta _{0}\left( T,{\bf v}_{s}\right) $) \cite{K0Sh(2003)} Eqs.
\ref{j(0)}-\ref{g(0)} can be rewritten to study the two components of the
current, $j_{x}$ (through the contact), and $j_{y}$ (along the contact):

\begin{equation}
g_{\omega }(0)=\frac{\widetilde{\omega }\Omega -i\frac{1}{2}sgn(v_{x})\Delta
_{0}^{2}\sin \phi }{\widetilde{\omega }^{2}+\Delta _{0}^{2}\cos ^{2}\frac{%
\phi }{2}}\equiv g_{y}+g_{x},  \label{g_s-wave}
\end{equation}
where we have split $g_{\omega }(0)$ in accordance with the two terms in the
numerator, which define $x$- and $y$-components of the current:

\begin{equation}
j_{i}(0)=16\pi eN_{0}v_{F}T%
\mathop{\rm Im}%
\sum_{\omega _{n}>0}\left\langle \widehat{v}_{i}g_{i}\right\rangle
_{v_{x}>0,v_{y}>0}.  \label{j_i}
\end{equation}
Here $\Omega =\sqrt{\widetilde{\omega }^{2}+\Delta _{0}^{2}}$, $i=x,y$. We
denote $j_{x}(0)\equiv j_{J}$ and detach transport supercurrent density $%
j_{T}$\ in $j_{y}(0)$, introducing $\widetilde{j}=j_{y}(0)-j_{T}$.

\begin{figure}[ph]
\centering
\includegraphics[width=8 cm]{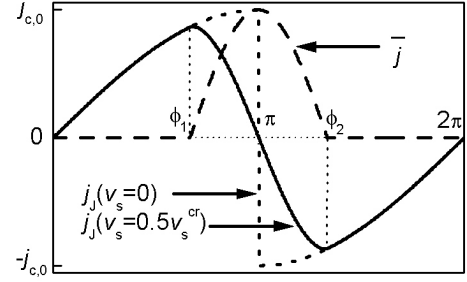}
\caption{Zero-temperature current-phase dependence for Josephson current $%
j_{J}$ (both in the absence and in the presence of the transport
supercurrent) and the countercurrent $\widetilde{j}$. The value of the
transport supercurrent is chosen to be half of the critical current (at $%
v_{s}=0.5\Delta _{0}/p_{F}$).}
\label{j_vs_fi}
\end{figure}

At $T=0$, from Eq.\ref{j_i} after integration\ we obtain (see also Fig.\ref
{j_vs_fi}) at $p_{F}v_{s}<\Delta _{\phi }$: $j_{J}=j_{J}^{(0)}=j_{c,0}sgn(%
\cos \frac{\phi }{2})\sin \frac{\phi }{2}$ and $\widetilde{j}=0$; at $\Delta
_{\phi }<p_{F}v_{s}<\Delta _{0}$:
\begin{equation}
j_{J}=j_{J}^{(0)}\left( 1-\frac{2}{\pi }\left( \arccos \beta -\beta \sqrt{%
1-\beta ^{2}}\right) \right) ,  \label{j_J_s-wave_1}
\end{equation}
\begin{equation}
\widetilde{j}=j_{c,0}\sin \frac{\phi }{2}\left( 1-\beta ^{2}\right) ,\ \beta
=\frac{\Delta _{\phi }}{p_{F}v_{s}},\ \Delta _{\phi }=\Delta _{0}\left| \cos
\frac{\phi }{2}\right| ,  \label{counter_j_s-wave}
\end{equation}
where $j_{c,0}=\frac{\pi }{2}\left| e\right| N_{0}v_{F}\Delta _{0}(T=0$,$%
v_{s}=0)$ is the critical Josephson current at $T=0$ and $v_{s}=0$; for $%
p_{F}v_{s}<\Delta _{0}$ the transport supercurrent is linear in $v_{s}$: $%
j_{T}=\frac{2}{3}eN_{0}v_{F}p_{F}v_{s}$. The condition $p_{F}v_{s}>\Delta
_{\phi }$ means that $\phi \in (\phi _{1},\phi _{2})$, where $\phi
_{1}=2\arccos \frac{p_{F}v_{s}}{\Delta _{0}}$, $\phi _{2}=2\pi -\phi _{1}$.

Thus, at $p_{F}v_{s}<\Delta _{\phi }$ the Josephson current is the same as
in the absence of the transport supercurrent and the current density along
the contact is equal to the transport supercurrent density. At $%
p_{F}v_{s}>\Delta _{\phi }$ for a fixed value of $\phi $, the Josephson
current $j_{J}$ is suppressed by the transport supercurrent (compared to the
Josephson current in the absence of the transport supercurrent): $%
j_{J}(v_{s})<j_{J}(v_{s}=0)$, and the countercurrent $\widetilde{j}$ appears
so that the total tangential current density at the contact $j_{y}(0)$
consists of the transport supercurrent $j_{T}$, carried by the condensate,
and of the countercurrent $\widetilde{j}$, carried by nonthermal
interface-induced quasiparticles. If there is a current at the contact which
flows in the direction opposite to the direction of the current far from the
contact (when $\widetilde{j}>j_{T}$), then the current distribution pattern
contains vortex-like formations (see also \cite{K0Sh(2003)}).

The dependence of the countercurrent $\left. \widetilde{j}(v_{s})\right|
_{\phi =\pi }$ on $v_{s}$\ at $T\neq 0$ can be deduced from the fact that
the current increases linearly with $v_{s}$ for $v_{s}<T/p_{F}$ and goes to
zero when $v_{s}$ tends to the critical value $v_{s}^{cr}(T)$ ($%
v_{s}^{cr}(0)\simeq p_{F}/\Delta _{0}$).

Our simple model allows straightforward generalizations. For example,
consider the situation when the boundary transparency is not one, $D\neq 1$%
\cite{RK04}. Then the proximity gap is $\Delta _{\phi }=\Delta _{0}\sqrt{%
1-D\sin ^{2}(\phi /2)}$ and the condition of the appearance of the
countercurrent (which is $p_{F}v_{s}>\Delta _{\phi }$ at $T=0$) at $D\neq 1$
is the stronger condition than the condition at $D=1$. At $T=0$ and with
given $v_{s}$ and $\phi $ the countercurrent appears for transparent enough
junctions with $D>D_{c}(v_{s},\phi ,T=0)=\left[ 1-\left( \frac{p_{F}v_{s}}{%
\Delta _{0}}\right) ^{2}\right] /\sin ^{2}\frac{\phi }{2}$.

Eq.\ref{g_s-wave} determines the energy of Andreev states: $\varepsilon
_{A}=\pm \Delta _{\phi }-{\bf p}_{F}{\bf v}_{s}$. Thus there are transport
current induced zero-energy states which are characterized by the values $%
\phi $ and $v_{s}$. At $T=0$ the conductance measures the excitation
spectrum of superconductor at the interface \cite{Duke}, \cite{FRS97}; the
zero-energy states are known to be responsible for the zero-bias conductance
peak (ZBCP) \cite{FRS97}, \cite{Hu}. At $\phi =0$, ZBCP is expected only in
the narrow interval of the value of the transport supercurrent (when there
are zero-energy states): $\Delta _{0}<p_{F}v_{s}<1.03\Delta _{0}$ \cite
{ZTH03}. In the system under consideration, where the proximity gap $\Delta
_{\phi }$\ is smaller than the bulk value $\Delta _{0}$, ZBCP is expected
even for smaller values of $v_{s}$: $\Delta _{\phi }<p_{F}v_{s}<1.03\Delta
_{0}$. The observation of ZBCP can be proposed as a test of
interface-induced transport-current-dependent quasiparticles states \cite
{FRS97}. Similar conclusions about a conductance peak due to the zero-energy
bound states were done in \cite{Amin} for the ac Josephson effect in the
point contact with phase difference $\phi $ close to $\pi $.

The appearance of the countercurrent can be understood as the response of
the weak link with negative self-inductance to the externally injected
transport supercurrent. The state of the junction in the absence of the
transport supercurrent at $T=0$\ is unstable at $\phi =\pi $\ from the point
of view that small deviations $\delta \phi =\pm 0$\ change $j_{J}$\ from $0$%
\ to $\mp j_{c,0}$. As was shown above, the response of the Josephson
junction to small transport supercurrent at $\phi =\pi $, $T=0$\ produces
the countercurrent $\widetilde{j}=j_{c,0}$. It is similar to the equilibrium
state with the persistent current in 1D normal metal ring with strong
spin-orbit interaction: there is degeneracy at $T=0$\ and $\phi =\pi $, and
the response of the ring is different at $\delta \phi \neq 0$\ or $B\neq 0$,
where $B$\ is the effective magnetic field which enters in the Hamiltonian
through the Zeeman term (which breaks time-reversal symmetry) \cite{1DNring}%
. The degeneracy is lifted by small effective magnetic field so that the
persistent current rapidly changes from $0$\ to its maximum value. In the
case of the weak link between two superconductors in the absence of the
transport supercurrent there is degeneracy between $+p_{y}$ and $-p_{y}$\
zero-energy states; both the time-reversal symmetry breaking by the surface
(interface) order-parameter and the Doppler shift (due to the transport
supercurrent or magnetic field) lift the degeneracy and result in the
surface (interface) current \cite{Walter},\cite{Ilichev}.

\section{Weak link between current-carrying {\it d}-wave superconductors}

\begin{figure}[ph]
\centering
\includegraphics[width=9 cm]{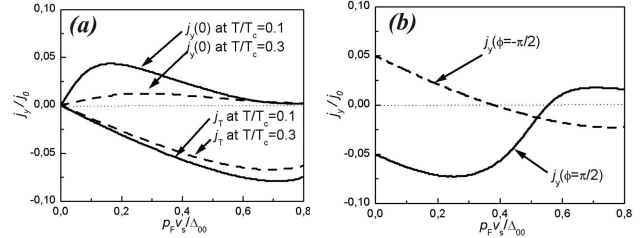}
\caption{(a) tangential current density $j_{y}$ in the banks, $%
j_{T}=j_{y}(\infty )$, and at the contact plane, $j_{y}(0)$, at $T=0.1T_{c}$
(solid lines) and $T=0.3T_{c}$ (dashed lines) versus superfluid velocity $%
v_{s}$; (b) tangential current density at the interface $j_{y}(0)$ for two
values of the phase difference versus $v_{s}$ ($T=0.1T_{c}$). Here $\Delta
_{00}=\Delta _{0}(T=0$,$v_{s}=0)$, $~j_{0}=4\protect\pi \left| e\right|
N(0)v_{F}T_{c}$.}
\label{j_vs_v_s}
\end{figure}

Here we consider the weak link between two {\it d}-wave superconductors
whose axes form an angle $\pi /4$ \cite{KOSh(2004)}. The link of this type
was proposed as a solid state qubit \cite{dd-qubit}. The equilibrium state
of the contact is doubly degenerate at $\phi =\pm \pi /2$; there is the
interface spontaneous current $j_{S}$\ in the equilibrium state; the
spontaneous currents flow in the opposite directions in the two states: $%
j_{S}\left( -\pi /2\right) =-j_{S}\left( \pi /2\right) $. One of the ways to
control the state of such a system is by means of transport supercurrent.
When the relative angle between two superconductors is equal to $\pi /4$,
the current at the contact is described by Eqs.\ref{j(0)}-\ref{g(0)} with $%
\Delta _{L}=\Delta _{0}\cos 2\vartheta $ and $\Delta _{R}=\Delta _{0}\sin
2\vartheta $ (here $\vartheta $\ is the azimuthal angle of ${\bf v}_{F}$).
The contact can be realized in a simply connected geometry, such as a grain
boundary, or in a doubly-connected geometry such as a ring (when the phase
difference is controlled by the magnetic flux piercing the ring).
Correspondingly we consider here two situations: $\phi =0$ (no external
phase difference) and $\phi =\pm \pi /2$ (in the ground state of the
contact). The effect of the transport supercurrent on the current parallel
to the boundary at $\phi =0$ is shown at Fig.\ref{j_vs_v_s}a.
Making use of Eqs.\ref{j(0)}-\ref{g(0)}, we plot the transport supercurrent
density $j_{T}$ far from the contact and the tangential component of the
current density at the contact $j_{y}(0)$. The latter flows in the opposite
direction to the transport supercurrent as in the case of contact of
conventional superconductors considered above. For the contact of {\it d}%
-wave superconductors, considered in this section, the countercurrent
appears for $\phi $ close to $0$ and $\pi $ (in contrast to the contact of
conventional superconductors, where it appears at $\phi \sim \pi $). This
countercurrent $\widetilde{j}=j_{y}(0)-j_{T}$, which is carried by the
quasiparticles, at low temperature rapidly grows at small $v_{s}$ as a
function of $v_{s}$ until it becomes equal to the critical value in
accordance with what was in detail described in \cite{Higashitani}. We also
note that at $T=0.3T_{c}$ our results are similar to the results of \cite
{FRS97}, presented there in Fig.2(a), where the dependence of the tangential
current density on the distance from the boundary for {\it d}-wave
superconductor was plotted. At this point our conclusions for the existence
of the countercurrent agree with the previous theoretical results \cite
{Higashitani},\cite{FRS97} which are consistent with experimental results
\cite{Braunisch}-\cite{Ilichev}. However, to the best of our knowledge,
there was no similar study for the case of the weak link (particularly,
between conventional superconductors), controlled by both $v_{s}$ and $\phi $%
, where the countercurrent along the boundary can appear, as described here.

At $\phi =\pm \pi /2$ (see Fig. \ref{j_vs_v_s}b) the transport supercurrent
removes the degeneracy when $j_{y}(0)=\pm j_{S}$. The dependence of $%
j_{y}(0) $ on $v_{s}$ is nonlinear. The resulting tangential current is not
always the sum of the spontaneous current $j_{S}$ and the transport
supercurrent $j_{T}$: $j_{y}(\phi =-\pi /2)\simeq j_{S}+j_{T}$; $j_{y}(\phi
=\pi /2)\simeq -j_{S}+j_{T}$ for $v_{s}<0.2\Delta /p_{F}$, i.e. until $%
\left| j_{y}\right| <j_{T}^{\max }$.

\section{Distribution of the current in the vicinity of the contact}

\begin{figure}[h]
\centering
\includegraphics[width=8.5 cm]{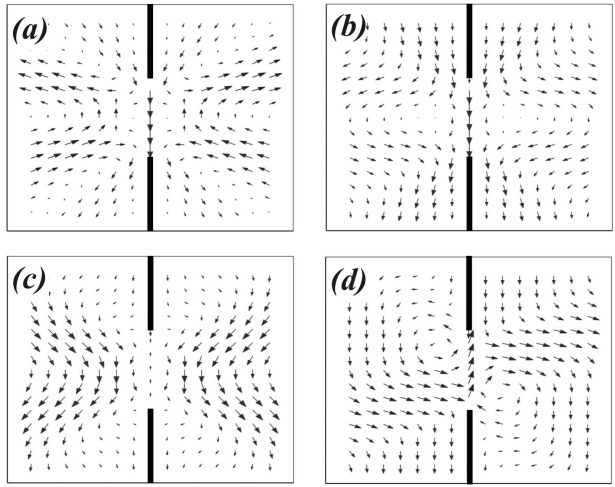}
\caption{Distribution of the current density in the vicinity of
the contact
for (a) $q=0$, $\protect\phi =\protect\pi /2$; (b) $q=0.2$ and $\protect\phi %
=\protect\pi /2$, (c) $q=0.2$ and $\protect\phi =3\protect\pi /2$; (d) $q=0.2
$ and $\protect\phi =\protect\pi $; here $q=p_{F}v_{s}/\Delta _{0}\left(
T=0,v_{s}=0\right) $.}
\label{DoD}
\end{figure}

To illustrate the spatial distribution of the current density in the
vicinity of the contact we study the case of ballistic point contact between
d-wave superconductors (see Fig. \ref{DoD}, where thick lines denote the
impenetrable partition between the superconducting banks). The current
distribution pattern in the vicinity of the point contact between
conventional superconductors was considered in \cite{K0Sh(2003)}.

Although the condition that the contact size is smaller than the
coherence length $a<\xi _{0}$\ is hardly realizable for
high-$T_{c}$\ superconductors, we consider this model as an
illustrative case to show: (a) how the current is distributed in
the ground state of the contact; (b)-(c) how the transport
supercurrent modifies the current distribution in the ground state
(qualitatively, the resulting current is a sum of what was in the
absence of $v_{s}$\ and of the transport current); (d) how the
appearance of the countercurrent results in the vortex-like
current distribution.

\section{Conclusion}

The weak link between current-carrying conventional and {\it d}-wave
superconductors has been studied. In such a system it is interesting to
study the coexistence of currents of different origin, both transport and
interference, or in other words, to understand how current-carrying states
of the banks mix non-locally in the vicinity of the contact. The system is
also interesting because of possible applications: in the Josephson
transistor with controlling parameters $\phi $ and $v_{s}$ governed by
external magnetic flux and the transport supercurrent \cite{Jtransistor},
and in solid-state qubits, based on a contact of {\it d}-wave
superconductors \cite{dd-qubit}. The current at the contact ({\it i.e.} its
components through the contact and along the contact plane) is controlled by
the values of $\phi $ and $v_{s}$. For a particular relation between $\phi $
and $v_{s}$ the countercurrent along the contact appears. This
countercurrent is carried by the interface-induced quasiparticles which
appear due to the coexistence of the transport supercurrent and the
proximity gap. When the countercurrent density is larger than the transport
supercurrent density there are vortex-like formations in the current
distribution in the vicinity of the contact.

\section{Acknowledgments}

The author is grateful to Yu.A. Kolesnichenko and A.N. Omelyanchouk for
their guidance and help during this research. The author also thanks E.V.
Bezuglyi for helpful discussions and P.K. Kovtun for careful reading of the
manuscript.


\begin{references}
\bibitem{Braunisch}  W. Braunisch {\it et al}., Phys. Rev. Lett. {\bf 68},
1908 (1992).

\bibitem{Walter}  H. Walter {\it et al}., Phys. Rev. Lett. {\bf 80}, 3598
(1998).

\bibitem{Ilichev}  E. Il'ichev {\it et al}., Phys. Rev. B {\bf 68}, 014510
(2003).

\bibitem{Higashitani}  S. Higashitani, J. Phys. Soc. Jpn. {\bf 66}, 2556
(1997).

\bibitem{FRS97}  M. Fogelstrom, D. Rainer, and J.A. Sauls, Phys. Rev. Lett.
{\bf 79}, 281 (1997).

\bibitem{KO}  I.O. Kulik and A.N. Omelyanchouk, Fiz. Nizk. Temp. {\bf 4, }%
296 (1978) (Sov. J. Low Temp. Phys. {\bf 4}, 142 (1978)).

\bibitem{K0Sh(2003)}  Yu.A. Kolesnichenko, A.N. Omelyanchouk, and S.N.
Shevchenko, Phys. Rev. B {\bf 67}, 172504 (2003).

\bibitem{KOSh(2004)}  Yu.A. Kolesnichenko, A.N. Omelyanchouk, and S.N.
Shevchenko, Fiz. Nizk. Temp. {\bf 30}, 288 (2004) (Low Temp. Phys. {\bf 30},
213 (2004)).

\bibitem{Jtransistor}  F.K. Wilhelm, G. Sh\"{o}n, and A.D. Zaikin, Phys.
Rev. Lett. {\bf 81}, 1682 (1998); E.V. Bezuglyi, V.S. Shumeiko, and G.
Wendin, preprint cond-mat/0303432.

\bibitem{dd-qubit}  L.B. Ioffe et al., Nature {\bf 398}, 679 (1999); A.\
Blais and A.M. Zagoskin, Phys. Rev. A {\bf 61}, 042308 (2000); A.M.
Zagoskin, J. Phys.: Condens. Matter {\bf 9}, L419 (1997).

\bibitem{RK04}  G. Rashedi and Yu.A. Kolesnichenko, Phys. Rev. B {\bf 69},
024516 (2004).

\bibitem{Duke}  C. Duke, ''Tunneling in solids'', Academic Press, NY (1969).

\bibitem{Hu}  C.-R. Hu, Phys. Rev. Lett. {\bf 72}, 1526 (1994).

\bibitem{ZTH03}  D. Zhang, C.S. Ting, and C.-R. Hu, preprint
cond-mat/0312545.

\bibitem{Amin}  M.H.S. Amin, Phys. Rev. B {\bf 68}, 054505 (2003).

\bibitem{1DNring}  T.-Z.Qian, Y.-S. Yi, and Z.-B. Su, Phys. Rev. B {\bf 55},
4065 (1997); V.A. Cherkassky, S.N. Shevchenko, A.S. Rozhavsky, I.D. Vagner,
Sov. J. Low Temp. Phys. {\bf 25}, 541 (1999); S.N. Shevchenko, Ph.D. thesis,
Inst. for Low Temp. Phys. \& Eng., Kharkov (2003).
\end{references}
\end{document}